\documentclass[aps,twocolumn,showpacs,preprintnumbers,amsmath,amssymb]{revtex4}
\usepackage{graphics}
\usepackage{epsfig}
\usepackage{dcolumn}% Align table columns on decimal point
\usepackage{bm}% bold math

\begin{document}
%\begin{CJK}{GBK}{song}
\title{The two-body open charm decays of $Z^+(4430)$}
\author{Xiang Liu$^{1,2}$}\email{liuxiang@teor.fis.uc.pt}
\author{Bo
Zhang$^3$}\email{olphice@163.com}
\author{Shi-Lin
Zhu$^1$}\email{zhusl@phy.pku.edu.cn} \affiliation{$^1$Department
of Physics, Peking University, Beijing 100871, China \\$^2$Centro
de F\'{i}sica Te\'{o}rica, Departamento de F\'{i}sica,
Universidade de Coimbra, P-3004-516, Coimbra, Portugal\\
$^3$Institute for Theoretical Physics, Regensburg University,
D-93040 Regensburg, Germany }

\date{\today}% It is always \today, today,

\begin{abstract}

The two-body open charm decays $Z^+(4430)\to D^{+}\bar{D}^{*0},
D^{*+}\bar{D}^{0}, D^{*+}\bar{D}^{*0}$ occur through the
re-scattering mechanism and their branching ratios are strongly
suppressed if $Z^+(4430)$ is a $D_1\bar D^*$ molecular state. In
contrast, $Z^+(4430)$ falls apart into these modes easily with
large phase space and they become the main decay modes if
$Z^+(4430)$ is a tetraquark state.  Experimental search of these
two-body open charm modes and the hidden charm mode
$\chi_{cJ}\rho$ will help distinguish different theoretical
schemes.

\end{abstract}


\maketitle

%%%%%%%%%%%%%%%%%%%%%%%%%%%%%%%%%%%%%%%%%%
\section{Introduction}
%%%%%%%%%%%%%%%%%%%%%%%%%%%%%%%%%%%%%%%%%%

Belle Collaboration observed a new resonance $Z^+(4430)$ in the
invariant mass spectrum of $\psi'\pi^+$ of $B\to K\psi'\pi^+$
\cite{Belle-4430}. Its mass and width are $m=4433\pm
4(\mathrm{stat})\pm 1(\mathrm{syst})$ MeV and
$\Gamma=44^{+17}_{-13}(\mathrm{stat})^{+30}_{-11}(\mathrm{syst})$
MeV respectively. This charged state stimulated extensive studies
of its nature
\cite{rosner,Meng,xiangliu,xiangliu-2,ding,qsr,maiani,cky,Gershtein,Bugg,Bugg2,Qiao,braaten,lu,close}.
Theoretical explanations of $Z^+(4430)$ include the S-wave
threshold effect \cite{rosner}, the $D_1(D_1')\bar D^*$ molecular
state \cite{Meng,xiangliu,xiangliu-2,ding,qsr}, the tetraquark
state \cite{maiani,cky,Gershtein}, the cusp effect \cite{Bugg} and
the $\Lambda_c-\Sigma_c^0$ bound state \cite{Qiao}.

Recently we studied whether $Z^+(4430)$ could be an S-wave
$D_1D^*$ ($D_1'D^*$) molecular state considering both the pion and
sigma meson exchange potentials \cite{xiangliu-2}. Our numerical
results show that there may exist an S-wave $D^*\bar{D}_1$
molecular state with $J^{P}=0^-$. If one ignores the width of
$D_1'$, there may also exist an S-wave $D^*\bar D_1'$ molecular
state with $J^{P}=0^-,1^-,2^-$. However, $D_1'$ will rapidly decay
into $D^*\pi$ before the formation of the $D_1'D^*$ bound state
because of its large width around 384 MeV.

In this work we assume $Z^+(4430)$ is a $D^*\bar{D}_1$ molecular
state with $J^{P}=0^-$ and study its two-body open charm decay
modes, which may be used to distinguish the tetraquark and
molecule picture. This work is organized as follows. After
introduction, we discuss the hidden {\sl vs} open charm decays of
$Z^+(4430)$. In Section \ref{amp}, the decay amplitude of the
two-body open charm decay of $Z^+(4430)$ is given. The last
section is the numerical results and discussion.

%%%%%%%%%%%%%%%%%%%%%%%%%%%%%%%%%%%%%%%%%%%%%%%%%%%%%%%%%%%%%%%%%%
\section{The hidden {\sl vs} open charm decays of $Z^+(4430)$}
%%%%%%%%%%%%%%%%%%%%%%%%%%%%%%%%%%%%%%%%%%%%%%%%%%%%%%%%%%%%%%%%%%

Assuming $Z^+(4430)$ is an S-wave ${D}^*\bar D_1$ molecular state,
we illustrate its strong decays allowed by Okubo-Zweig-Iizuka
(OZI) rule in Fig. \ref{haha}. Here indices 1, 2, 3, and 4 are
$c$, $\bar{u}$, $\bar{c}$ and $d$ quarks respectively. The
dashed-line box denotes the S-wave ${D}^*-\bar D_1$ molecular
state.

\subsection{Hidden charm decays}

\begin{center}\begin{figure}[htb]
\begin{tabular}{c}
\scalebox{0.8}{\includegraphics{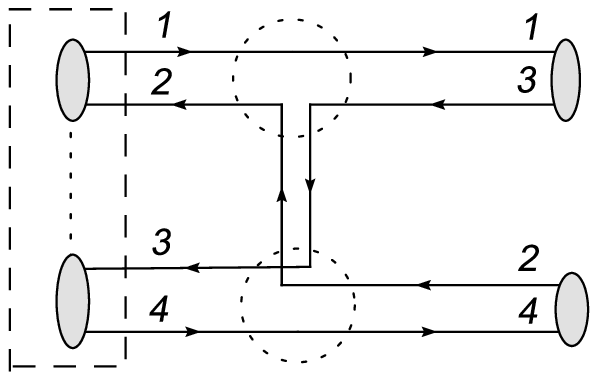}}\\\\
(a)\\\\ \scalebox{0.46}{\includegraphics{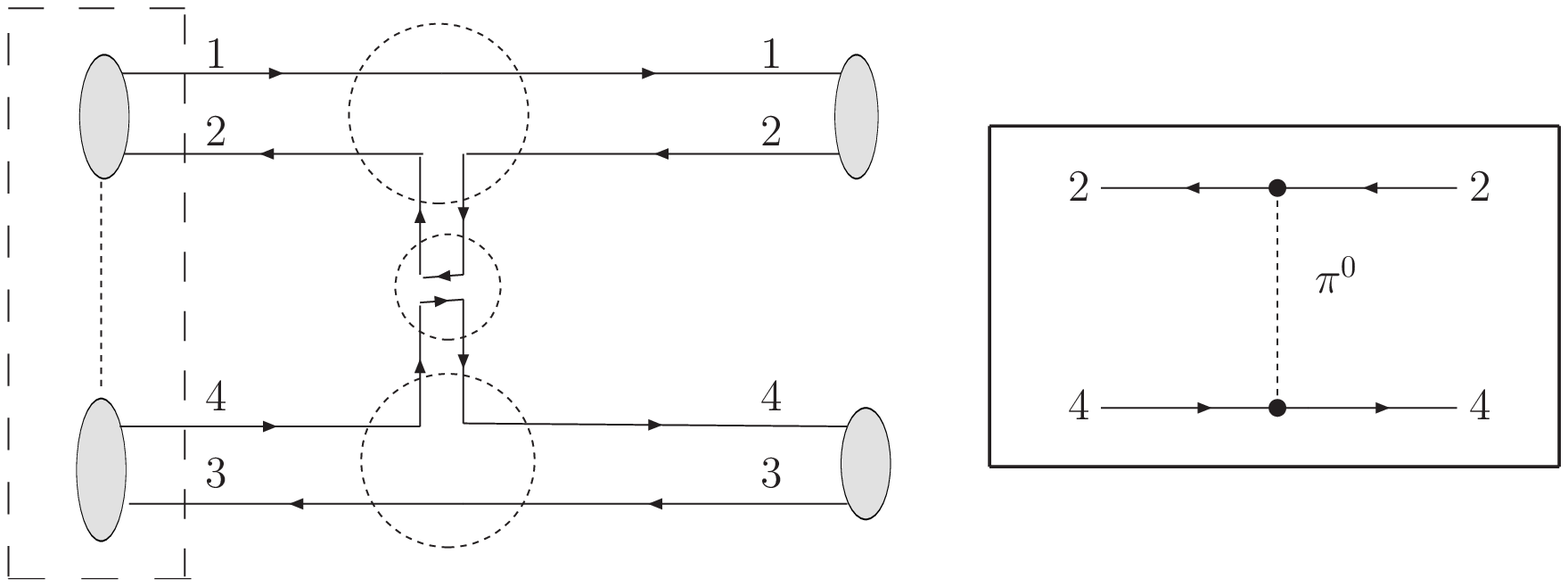}}\\\\
(b)
\end{tabular} \caption{Diagrams (a) and (b) depict the hidden
and open charm decay of $Z^+(4430)$ at the quark level
respectively . Here 1, 2, 3, and 4 denote $c$, $\bar{u}$,
$\bar{c}$ and $d$ quarks respectively . \label{haha}}
\end{figure}\end{center}

Fig. \ref{haha} (a) describes the hidden charm decay process
$Z^+(4430)\to {D}^*\bar D_1 \to (c\bar{c})(d\bar{u})$ by
exchanging one charmed meson between ${D}^*$ and $\bar D_1$.
$Z^+(4430)$ was observed in the $\psi'\pi^+$ channel. If it's the
S-wave ${D}^*\bar D_1$ molecular state, its quantum number is
$I^{G}(J^{P})=1^+(0^-,1^-,2^-)$ \cite{xiangliu}. Kinematically
allowed hidden charm decays can be classified into (1) P-wave
modes: $\psi\pi$, $\psi'\pi$, $\psi(3S)\pi$, $\psi(4S)\pi$,
$\psi(1D)\pi$, $\psi(2D)\pi$, $\eta_c\rho$, $\eta_c(2S)\rho$; (2)
S-wave modes: $\chi_{cJ}\rho$. The neutral partner of $Z^+(4430)$
may also decay into $\psi\omega$ via P-wave and $\chi_{cJ}\eta$
via S-wave.

$Z^+(4430)$ was observed only in the $\psi'\pi^+$ channel up to
now, which is very puzzling. One of the possible reasons is the
mismatch of the Q-values of the initial and final states
\cite{xiangliu}. If so, one should also expect a signal in the
$\pi^+\psi(3S)$ channel since there is nearly no mismatch of the
Q-value. Another potential reason is the specific nodal structure
in the wave functions of the final states, which was discussed by
Bugg recently \cite{Bugg2}. Meng and Chao adopted the
re-scattering mechanism to explain the suppression of the
$\pi^+J/\psi$ decay mode \cite{Meng}.

Besides the discovery mode $\psi'\pi^+$, we want to emphasize that
the S-wave $\chi_{cJ}\rho^+$ mode is very important since there is
no additional suppression factor $({k\over m_Z})^2$ where $k$ is
the decay momentum and $m_Z$ is the mass of $Z^+(4430)$. Naively
one may expect that the orbital excitation in the $D_1$ meson
easily transfers to $\chi_{cJ}$. Moreover, the $\chi_{c0}\rho^+$
mode may help distinguish the $J^{P}$ of $Z^+(4430)$ and test the
${D}^*\bar D_1$ molecular state assignment. If the
$\chi_{c0}\rho^+$ mode is observed, one can exclude the $J^P=0^-$
possibility for the ${D}^*\bar D_1$ system. We strongly urge
experimental colleagues to search $Z^+(4430)$ in the
$\chi_{cJ}\pi^+\pi^0$ channel.

\subsection{Open charm decays}

Both $Z^+(4430)$ and $D_1$ mesons have finite widths. $Z^+(4430)$
can decay through the upper tail of its mass distribution into
$D_1\bar D^\ast$, especially the lower part of the $D_1$ mass
distribution. As pointed out in Ref. \cite{Meng}, the dominant
decay mode of $Z^+(4430)$ is $Z^+(4430)\to D_1 \bar D^\ast \to
D^\ast \bar D^\ast \pi$.

Besides the above dominant mode, there exists another class of
open charm decay mode through the re-scattering mechanism
\cite{rescattering,chao}, which is shown in Fig. \ref{haha} (b).
One pion or $\rho$ ($\omega$) meson exchange occurs between the
light quarks. These two-body decay modes include $D\bar{D}^*
+D^*\bar{D}$ and $D^*\bar{D}^*$. The $D\bar D$ decay mode is
forbidden for a pseudoscalar meson.

This kind of open charm decay is strongly suppressed if $Z(4430)$
is a molecular state, as will be shown in the following sections.
However, $Z(4430)$ will fall apart into these two-body final
states very easily with large phase space if $Z(4430)$ is a
tetraquark state as proposed in Refs. \cite{maiani,cky,Gershtein}.
Therefore, these two-body decay modes can be used to distinguish
the molecular and tetraquark pictures. This is the key motivation
of the present work.

%%%%%%%%%%%%%%%%%%%%%%%%%%%%%%%%%%%%%%%%%%%%%%%%%%%%%%%%%%%%%%%%%%%%
\section{The amplitude of the two-body open charm decay}\label{amp}
%%%%%%%%%%%%%%%%%%%%%%%%%%%%%%%%%%%%%%%%%%%%%%%%%%%%%%%%%%%%%%%%%%%%

Assuming $Z^+(4430)$ is an S-wave ${D}^*-\bar D_1$ molecular state
with $J^P=0^-$ \cite{xiangliu-2}, we derive the two-body open
charm decay amplitude using the optical theorem. From the
unitarity of the S-matrix $S^{\dagger}S=1$, one obtains
\begin{eqnarray}
-i(T-T^{\dagger})=T^{\dagger}T
\end{eqnarray}
where $S=1+iT$. We sandwich the above equation between the
one-particle state $|P\rangle$ and two-particle state
$|k_1,k_2\rangle$. By inserting a complete set of intermediate
states to the right-hand side, one gets
\begin{eqnarray}
&&\langle
k_1,k_2|T^{\dagger}T|P\rangle\nonumber\\&=&\sum_{\alpha}(\prod_{i=1}^{\alpha}\int\frac{d^3
p_i}{(2\pi)^3}\frac{1}{2E_i})\langle
k_1,k_2|T^{\dagger}|p_i\rangle\langle p_i|T|P\rangle.
\end{eqnarray}
Now the absorptive part reads
\begin{eqnarray}
&&\mathtt{Abs}\; \mathcal{M}(A(P)\to B(k_1)
C(k_2))\nonumber\\&=&\frac{1}{2}\sum_{\alpha}\Big(\prod_{i=1}^{\alpha}\int\frac{d^3
p_i}{(2\pi)^3}\frac{1}{2E_i}\Big)(2\pi)^4\delta^4(k_1+k_2-\sum_{i=1}^{\alpha}
p_i)\nonumber\\&&\times \mathcal{M}[k_1k_2\to
\{p_{i}\}]\mathcal{M}^*[P\to \{p_{i}\}].\label{abs}
\end{eqnarray}
The optical theorem relates the absorptive part of the $A\to B+C$
decay amplitude to the sum of all possible $A\to\{p_i\}$ decays
and the re-scattering process $\{p_i\}\to B+C$.

Thus, for the two-body charm decay $Z^+(4430)\to \bar{D}_1D^*\to
D^{(*)}\bar{D}^{(*)}$, the absorptive part of the decay amplitude
can be written as
\begin{eqnarray}
&&\mathtt{Abs}[Z^+(4430)\to \bar{D}_1D^*\to D^{(*)}\bar{D}^{(*)}]
\nonumber\\&=&\frac{|\mathbf{p}|}{32\pi^2 M_Z}\int d\Omega
\mathcal{M}^*[Z^+\to D^*\bar{D}_1]\nonumber\\&&\times\mathcal{M}[
D^*\bar{D}_1\to D^{(*)}\bar D^{(*)}]
\end{eqnarray}
where $|\bf p|$ denotes the three momentum of intermediate state
in the center of mass frame of $Z^+(4430)$.

Because the mass of $Z^+(4430)$ is close to the sum of the masses
of $D_1$ and $D^*$, the dispersive part of $Z^+(4430)\to
\bar{D}_1D^*\to D^{(*)}\bar{D}^{(*)}$ amplitude is very important
as shown in the case of X(3872) by Meng and Chao \cite{chao}. The
dispersive part of the amplitude is related to the absorptive part
through the dispersion relation,
\begin{eqnarray}
\mathtt{Dis}
\mathcal{M}(M_Z)=\frac{1}{\pi}\int^{\infty}_{s_{0}^2}\frac{\mathtt{Abs}
\mathcal{M}(s)}{s-M_{Z}^2}ds,
\end{eqnarray}
with
\begin{eqnarray}
\mathtt{Abs} \mathcal{M}(s)&=&\mathtt{Abs}[Z^+(4430)\to
\bar{D}_1D^*\to D^{(*)}\bar{D}^{(*)}]\nonumber\\&&\times\exp(-\eta
|\mathbf{k}|^2),
\end{eqnarray}
where
$|\mathbf{k}|=[\lambda(M_Z^2,m_{D^*}^2,m_{D_1}^2)]^{1/2}/2M_Z$ is
the three momentum of the intermediate state in the rest frame of
$Z^+(4430)$. $\lambda(a,b,c)=a^2+b^2+c^2-2ab-2ac-2bc$ is
k\"{a}llen function. One chooses $s_0=m_{D_1}+m_{D^*}$. The
exponential not only describes the dependence of the interaction
between $Z^+(4430)$ and $D^*\bar{D}_1$ on $|\mathbf{k}|$, but also
plays the role of the cutoff. The factor $\eta$ is related to the
interaction radius $R$ by $\eta = R/6$ \cite{Pennington}.

In order to get $\mathcal{M}^*[Z^+\to D^*\bar{D}_1]$ and
$\mathcal{M}[ D^*\bar{D}_1\to D^{(*)}D^{(*)}]$, we use the
following Lagrangians \cite{Meng,Lagrangian}
%\begin{widetext}
\begin{eqnarray}
\mathcal{L}_{Z^+D_1 D^*}&=&g_{Z^+D_1 D^*}Z(D^*\cdot
D_1^{\dagger})+h.c.,
\end{eqnarray}
\begin{eqnarray}
\mathcal{L}_1&=&\frac{1}{2}g_{_{\mathcal{D^{*}}\mathcal{D^{*}}\mathbb{P}}}\varepsilon_
{\mu\nu\alpha\beta}\mathcal{D^* }_{i}^{\mu}
\partial^{\nu}\mathbb{P}^{ij}{\stackrel{\leftrightarrow}{\partial}}^
{\alpha}\mathcal{D^{*}}_{j}^{\beta\dagger}\nonumber\\&&-ig_{_{\mathcal{D^{*}}\mathcal{D}\mathbb{P}}}(\mathcal{D}^{i}\partial^{\mu}\mathbb{P}_{ij}\mathcal{D^{*}}_{\mu}^{j\dagger}
-\mathcal{D^{*}}_{\mu}^{i}\partial
^{\mu}\mathbb{P}_{ij}\mathcal{D}^{j\dagger})\nonumber\\
&&-ig_{_{\mathcal{D}\mathcal{D}\mathbb{V}}}\mathcal{D}_{i}^{\dagger}{\stackrel{\leftrightarrow}{\partial}}
_{\mu}\mathcal{D}^{j}(\mathbb{V}^{\mu})^{i}_{j}
\nonumber\\&&-2f_{_{\mathcal{D^{*}}\mathcal{D}\mathbb{V}}}\varepsilon_{\mu\nu\alpha\beta}(\partial^{\mu}\mathbb{V}^{\nu})
^{i}_{j}(\mathcal{D}_{i}^{\dagger}
{\stackrel{\leftrightarrow}{\partial}}^{\alpha}\mathcal{D^{*}}^{\beta
j}-\mathcal{D^{*}}_{i}^{\beta
\dagger}{\stackrel{\leftrightarrow}{\partial}}^{\alpha}\mathcal{D}^{j})\nonumber\\
&&+ig_{_{\mathcal{D^{*}}\mathcal{D^{*}}\mathbb{V}}}\mathcal{D^{*}}_{i}^{\nu
\dagger}{\stackrel{\leftrightarrow}{\partial}}_{\mu}\mathcal{D^{*}}_{\nu}^{j}(\mathbb{V}^{\mu})^{i}_{j}\nonumber\\&&+
4if_{_{\mathcal{D^{*}}\mathcal{D^{*}}\mathbb{V}}}\mathcal{D^{*}}_{i\mu}^{\dagger}(\partial^{\mu}\mathbb{V}^{\nu}-\partial^{\nu}
\mathbb{V}^{\mu})^{i}_{j}
\mathcal{D^{*}}_{\nu}^{j},\label{lagrangian}
\end{eqnarray}
\begin{eqnarray}
\mathcal{L}_{D_1D^*\pi}&=&ig_{D_{1}D^*\pi}[-3
D^{*}_{\mu}D_{1\nu}^{\dagger}\partial^{\mu}\partial^{\nu}\pi+(D^*\cdot
D_1^{\dagger})\partial^2\pi\nonumber\\&&-\frac{1}{m_{D^*}m_{D_1}}\partial_{\mu}D^{*\rho}\partial_{\nu}D_{1\rho}^\dagger
\partial^{\mu}\partial^{\nu}\pi]+h.c.,
\end{eqnarray}
%\end{widetext}
where $\mathcal{D}$ and $\mathcal{D^*}$ are pseudoscalar and
vector heavy mesons respectively, i.e.
$\mathcal{D^{(*)}}$=(($\bar{D}^{0})^{(*)}$, $(D^{-})^{(*)}$,
$(D_{s}^{-})^{(*)}$). $\mathbb{P}$ and $\mathbb{V}$ denote the
octet pseudoscalar and the nonet vector meson matrices. The values
of the coupling constants will be given in Section
\ref{numerical}. The Lagrangian relevant to the interaction of
$D_1$ with $\rho(\omega)$ and $D^{(*)}$ mesons is given in Ref.
\cite{zhu-QSR}.

The flavor wave function of $Z^+(4430)$ is \cite{xiangliu}
\begin{equation}
|Z^+(4430)\rangle
=\frac{1}{\sqrt{2}}\Big(|D^{*+}\bar{D_1}^0\rangle+|\bar{D}^{*0}{D_1}^+\rangle\Big)\;
.
\end{equation}
Thus $Z^+(4430)$ with $J^P=0^-$ can decay into $D^{*+}\bar{D}^0$,
$D^+\bar{D}^{*0}$ and $D^{*+}\bar D^{*0}$ via the intermediate
states $D^{*+}\bar{D_1}^0$ and $\bar{D}^{*0}{D_1}^+$. If one
ignores the mass difference between neutral and charged charmed
mesons, the contribution from the intermediate state
$D^{*+}\bar{D_1}^0$ is same as that from $\bar{D}^{*0}{D_1}^+$.

\subsection{$Z^+(4430)\to D^{*+}\bar{D_1}^0(\bar{D}^{*0}{D_1}^+)\to
\bar{D}^{*0}D^+$}\label{B}

\begin{center}\begin{figure}[htb]
\begin{tabular}{cc}
\scalebox{0.42}{\includegraphics{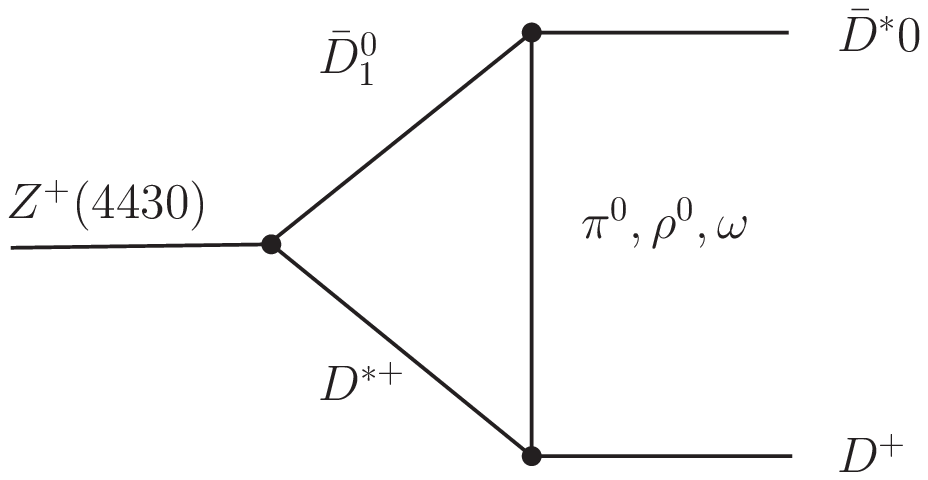}}&\scalebox{0.42}{\includegraphics{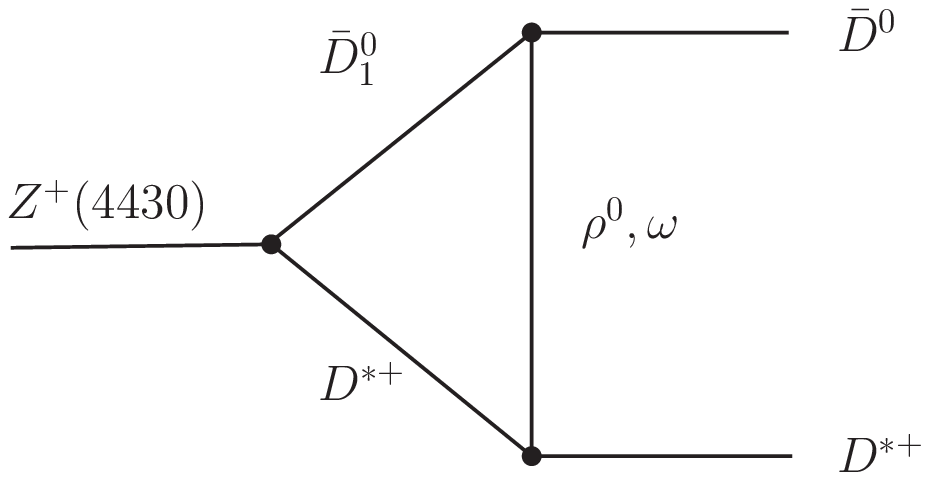}}\\\\
(a)&(b)\\\\ \scalebox{0.42}{\includegraphics{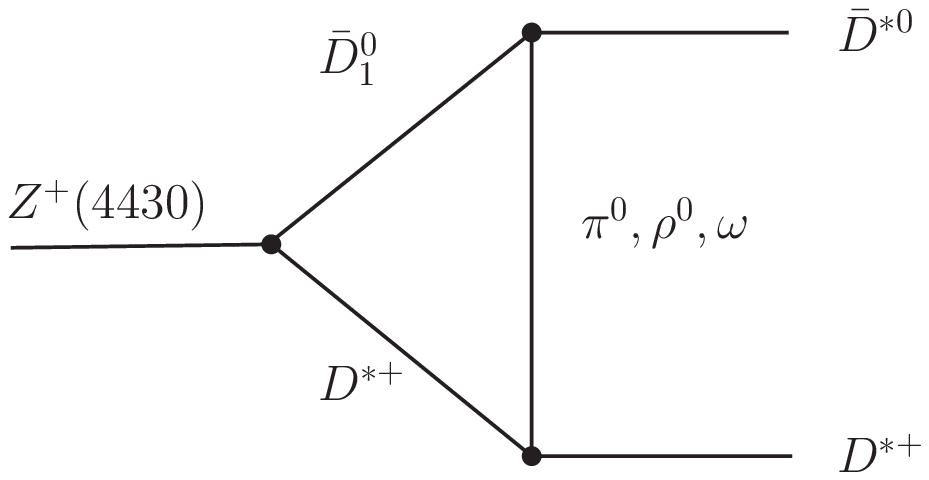}}&\\\\
(c)&
\end{tabular} \caption{The two-body open charm decays of $Z^+(4430)$. \label{haha1}}
\end{figure}\end{center}

Fig. \ref{haha1} (b) describes the open charm decay $Z^+(4430)\to
D^{*+}\bar{D_1}^0\to \bar{D}^{*0}D^+$. The absorptive part of the
decay amplitude of $Z^+(4430)\to
\bar{D}_1^{0}(p_1,\epsilon_1)D^{*+}(p_2,\epsilon_2)\to
\bar{D}^{*0}(p_3,\epsilon_3)D^+(p_4)$ via the $\pi$ meson exchange
is
\begin{eqnarray}
&&\mathtt{Abs}^{\pi}[Z^+\to D^{*+}\bar{D_1}^0\to
D^+\bar{D}^{*0}]\nonumber\\&=&\frac{|\mathbf{p}|}{32\pi^2 M_Z}\int
d\Omega \;i  [i g_{Z^+D_1 D^*}%\underline{\epsilon_1\cdot
%\epsilon_2}
]\nonumber\\&&\times[i(-i)\frac{g_{D^*DP}}{\sqrt
{2}}(iq^{\mu})%\underline{\epsilon_{2\mu}}
]
ig_{D_1D^*\pi}[-3q^{\beta}q^{\nu}\nonumber\\&&+g^{\beta\nu}q^2-\frac{g^{\nu\beta}}{m_{D^*}m_{D_1}}(p_3\cdot
q)(p_1\cdot
q)]\epsilon_{3\beta}%\underline{\epsilon_{1\nu}}
\nonumber\\&&\times
\Big[-g_{\nu\rho}+\frac{p_{1\nu}p_{1\rho}}{m_{D_1}^2}\Big]\Big[-g_{\mu}^{\rho}+\frac{p_{2}^{\rho}p_{2\mu}}{m_{D^*}^2}\Big]
\nonumber\\&&\times\frac{\mathcal{F}^2(q^2,m_i^2)}{q^2-m_\pi^2}.
\end{eqnarray}

The amplitude from the $\rho(\omega)$ exchange is
\begin{eqnarray}
&&\mathtt{Abs}^{\rho(\omega)}[Z^+\to D^{*+}\bar{D_1}^0\to
D^+\bar{D}^{*0}]\nonumber\\&=&\frac{|\mathbf{p}|}{32\pi^2 M_Z}\int
d\Omega \; i [i g_{Z^+D_1 D^*}%\underline{\epsilon_1\cdot
%\epsilon_2}
]\nonumber\\&&\times[-2I_1^{(i)}i
f_{D^*DV}\varepsilon_{\mu\nu\alpha\beta}(iq^{\mu})%\underline{\epsilon_{\rho}^{\nu}}
(-ip_4^{\alpha}-ip_2^{\alpha})
%\underline{\epsilon_2^{\beta}}
]\nonumber\\&&\times
I_2^{(i)}%\underline{\epsilon_{\rho}^{\tau}}%\underline{\epsilon_{1}^{\eta}}
[i\varepsilon_{\tau\sigma\xi\kappa}
\frac{p_1^{\kappa}}{m_{D_1}}\epsilon_{3}^{\xi}\Theta^{\sigma}_{\eta}(g_{2,2}'+g_{1,2}')
\nonumber\\&&+i\varepsilon_{\tau\sigma\eta\kappa}
\frac{p_1^{\kappa}}{m_{D_1}}\epsilon_{3\gamma}\Theta^{\sigma\gamma}(g_{2,2}'-g_{1,2}')
+i\varepsilon_{\tau\eta\xi\kappa}\epsilon_{3}^{\xi}\frac{p_1^{\kappa}}{m_{D_1}}g_{1,0}']\nonumber\\&&
\times\Big[-g^{\tau\nu}+\frac{q^{\tau}q^{\nu}}{m_{i}^2}\Big]\Big[-g^{\eta}_{\lambda}+\frac{p_1^{\eta}p_{1\lambda}}{m_{D_1}^2}\Big]
\nonumber\\&&\times\Big[-g^{\beta\lambda}+\frac{p_{2}^{\beta}p_{2}^{\lambda}}{m_{D^*}^2}\Big]
\frac{\mathcal{F}^2(q^2,m_i^2)}{q^2-m_i^2}
\end{eqnarray}
with
\begin{eqnarray}
&&\Theta^{ab}\nonumber\\&=&\frac{g^{ab}(q\cdot
p_1)^2}{3m_{D_1}^2}+\frac{2p_1^{a}p_1^{b}(q\cdot
p_1)^2}{3m_{D_1}^4}-\frac{p_1^{a}q^{b}(q\cdot
p_1)}{m_{D_1}^2}\nonumber\\&&-\frac{p_1^{b}q^{a}(q\cdot
p_1)}{m_{D_1}^2}+q^a q^b-\frac{g^{ab}q^2}{3}+\frac{p_1^a p_1^b
q^2}{3m_{D_1}^2},
\end{eqnarray}
where the index $i$ in $I_{1,2}^{(i)}$ and $m_{i}$ denotes the
$\rho(\omega)$ meson.
$I_1^{(\rho)}=-I^{(\omega)}_2=-\frac{1}{\sqrt{2}}$ and
$I_{2}^{(\rho)}=I_2^{(\omega)}=\frac{1}{\sqrt{2}}$. The same
notations are also used in the following subsections.

\subsection{$Z^+(4430)\to D^{*+}\bar{D_1}^0(\bar{D}^{*0}{D_1}^+)\to
\bar{D}^0{D}^{*+}$}\label{C}

The open charm decay $Z^+(4430)\to D^{*+}\bar{D_1}^0\to
\bar{D}^{0}D^{*+}$ is depicted in Fig. \ref{haha1} (c). The
absorptive part of the decay amplitude of $Z^+(4430)\to
\bar{D}_1^{0}(p_1,\epsilon_1)D^{*+}(p_2,\epsilon_2)\to
\bar{D}^{0}(p_3)D^{*+}(p_4,\epsilon_4)$ from the $\rho(\omega)$
meson exchange is
\begin{eqnarray}
&&\mathtt{Abs}^{\rho(\omega)}[Z^+\to D^{*+}\bar{D_1}^0\to
D^{*+}\bar{D}^0]\nonumber\\&=&\frac{|\mathbf{p}|}{32\pi^2 M_Z}\int
d\Omega \;i  [i g_{Z^+D_1 D^*}%\underline{\epsilon_1\cdot
%\epsilon_2}
]\nonumber\\&&\times
I_{1}^{(i)}[-2ig_{D^*D^*V}p_{2\nu}\epsilon_{4}^{\mu}-4if_{D^*D^*V}(q^{\mu}\epsilon_{4\nu}-g_{\nu}^{\mu}
q\cdot\epsilon_4)]%\underline{\epsilon_{\rho}^{\nu}\epsilon_{2\mu}}]
\nonumber\\&&\times
I_{2}^{(i)}%\underline{\epsilon_{1\kappa}}\underline{\epsilon_{\rho\xi}}
\Big[ig_{2,2}
\Theta^{\kappa\xi}+ig_{1,0}(g^{\kappa\xi}-\frac{p_1^{\kappa}p_1^{\xi}}{m_{D_1}^2})\Big]\Big[-g^{\nu}_{\xi}+\frac{q_{\xi}q^{\nu}}{m_{i}^2}\Big]\nonumber\\&&
\times\Big[-g^{\sigma}_{\kappa}+\frac{p_1^{\sigma}p_{1\kappa}}{m_{D_1}^2}\Big]
\Big[-g_{\sigma\mu}+\frac{p_{2\sigma}p_{2\mu}}{m_{D^*}^2}\Big]\frac{\mathcal{F}^2(q^2,m_i^2)}{q^2-m_i^2}.\nonumber\\
\end{eqnarray}

\subsection{$Z^+(4430)\to D^{*+}\bar{D_1}^0(\bar{D}^{*0}{D_1}^+)\to
D^{*+}\bar{D}^{*0}$} \label{D}

Fig. \ref{haha1} (d) corresponds to the process $Z^+(4430)\to
\bar{D_1}^0(p_1,\epsilon_1)D^{*+}(p_2,\epsilon_2)\to
\bar{D}^{*0}(p_3,\epsilon_3)D^{*+}(p_4,\epsilon_4)$. The
absorptive part of the amplitude from the $\pi$ exchange is
\begin{eqnarray}
&&\mathtt{Abs}^{\pi}[Z^+\to D^{*+}\bar{D_1}^0\to
D^{*+}\bar{D}^{*0}]\nonumber\\&=&\frac{|\mathbf{p}|}{32\pi^2
M_Z}\int
d\Omega \;i  [i g_{Z^+D_1 D^*}%\underline{\epsilon_1\cdot
%\epsilon_2}
]\nonumber\\&&\times[-\frac{ig_{D^*D^*P}}{2\sqrt
{2}}\varepsilon_{\mu\tau\alpha\xi}(iq^{\tau})%\underline{\epsilon_{2}^{\mu}}
(ip_2^{\alpha}+ip_4^{\alpha})\epsilon_4^{\xi}]
ig_{D_1D^*\pi}\nonumber\\&&\times[-3q^{\beta}q^{\nu}+g^{\beta\nu}q^2-\frac{g^{\nu\beta}}{m_{D^*}m_{D_1}}(p_3\cdot
q)(p_1\cdot
q)]\epsilon_{3\beta}%\underline{\epsilon_{1\nu}}
\nonumber\\&&\times
\Big[-g_{\nu\rho}+\frac{p_{1\nu}p_{1\rho}}{m_{D_1}^2}\Big]\Big[-g^{\mu\rho}+\frac{p_{2}^{\rho}p_{2}^{\mu}}{m_{D^*}^2}\Big]
\frac{\mathcal{F}^2(q^2,m_i^2)}{q^2-m_\pi^2}.\nonumber\\
\end{eqnarray}
For the $\rho$ meson exchange, the amplitude is
\begin{eqnarray}
&&\mathtt{Abs}^{\rho(\omega)}[Z^+\to D^{*+}\bar{D_1}^0\to
D^{*+}\bar{D}^{*0}]\nonumber\\&=&\frac{|\mathbf{p}|}{32\pi^2
M_Z}\int
d\Omega \;i  [i g_{Z^+D_1 D^*}%\underline{\epsilon_1\cdot
%\epsilon_2}
]\nonumber\\&&\times
I_{1}^{(i)}[-2ig_{D^*D^*V}p_{2\nu}\epsilon_{4}^{\mu}-4if_{D^*D^*V}(q^{\mu}\epsilon_{4\nu}-g_{\nu}^{\mu}
q\cdot\epsilon_4)]%\underline{\epsilon_{\rho}^{\nu}\epsilon_{2\mu}}
%\underline{\epsilon_2^{\beta}}
\nonumber\\&&\times
I_2^{(i)}%\underline{\epsilon_{\rho}^{\tau}}\underline{\epsilon_{1}^{\eta}}
[i\varepsilon_{\tau\sigma\xi\kappa}
\frac{p_1^{\kappa}}{m_{D_1}}\epsilon_{3}^{\xi}\Theta^{\sigma}_{\eta}(g_{2,2}'+g_{1,2}')
\nonumber\\&&+i\varepsilon_{\tau\sigma\eta\kappa}
\frac{p_1^{\kappa}}{m_{D_1}}\epsilon_{3\gamma}\Theta^{\sigma\gamma}(g_{2,2}'-g_{1,2}')
+i\varepsilon_{\tau\eta\xi\kappa}\epsilon_{3}^{\xi}\frac{p_1^{\kappa}}{m_{D_1}}g_{1,0}']\nonumber\\&&
\times\Big[-g^{\nu\tau}+\frac{q^{\tau}q^{\nu}}{m_{i}^2}\Big]\Big[-g^{\sigma\eta}+\frac{p_1^{\sigma}p_{1}^{\eta}}{m_{D_1}^2}\Big]
\Big[-g_{\sigma\mu}+\frac{p_{2\sigma}p_{2\mu}}{m_{D^*}^2}\Big]\nonumber\\&&\times\frac{\mathcal{F}^2(q^2,m_i^2)}{q^2-m_i^2}.
\end{eqnarray}

Throughout this section, $\mathcal{F}(q^2,m_i^2)$ is the form
factor to describe the structure effect in the vertex of the
re-scattering process $\bar{D}_{1}D^*(D_1\bar{D}^*)\to
D^{(*)}\bar{D}^{(*)}$. We use the monopole form
\cite{hycheng,rescattering}
\begin{eqnarray}
\mathcal{F}(q^2,m_i^2)=\Big(\frac{m_i^2-\Lambda^2}{q^2-\Lambda^2}\Big)
\end{eqnarray}
where $\Lambda=m_{i}+\alpha \Lambda_{QCD}$ with
$\Lambda_{QCD}=220$ MeV and $\alpha=1\sim 3$.

%%%%%%%%%%%%%%%%%%%%%%%%%%%%%%%%%%%%%%%%%%%%%%%%%%%%%%%%%%%
\section{Results and discussion}\label{numerical}
%%%%%%%%%%%%%%%%%%%%%%%%%%%%%%%%%%%%%%%%%%%%%%%%%%%%%%%%%%%

We collect the values of these coupling constants below
\begin{eqnarray*}
g_{D^*D^*P}&=&\frac{g_{D^*DP}}{\sqrt{m_{D}m_{D^*}}}=\frac{2g}{f_{\pi}},\\
f_{D^*DV}&=&\frac{f_{D^*D^*V}}{m_{D^*}}=\frac{\lambda g_{V}}{\sqrt{2}},\;\;g_{D^*D^*V}=\frac{\beta g_{V}}{\sqrt{2}},\\
g_{D_1
D^*\pi}&=&-\frac{\sqrt{m_{D^*}m_{D_1}}}{3f_{\pi}\Lambda_{\chi}}(h_1+h_2),\;\;
g_{V}=\frac{m_{\rho}}{f_{\pi}},
\end{eqnarray*}
where $g=0.59$, $\lambda=0.56$ GeV$^{-1}$, $\beta=0.9$ are the
parameters in the effective Lagrangian
\cite{Lagrangian,hycheng,Cleo,Isola}. Casalbuoni and his
collaborators extracted $h'=(h_1+h_2)/\Lambda_{\chi}=0.55$
GeV$^{-1}$ with the available experimental data \cite{Lagrangian}.
In Ref. \cite{zhu-QSR}, Zhu and Dai gave $g_{2,2}=\sqrt{6}g_d/6 $,
$g_{1,0}=-\sqrt{6}g_s/3 $, $g_{2,2}'=\sqrt{6}g_d/4$,
$g_{1,2}'=\sqrt{6}g_d/12$, $g_{1,0}'=\sqrt{6}g_s/6 $, $g_s=2.1$
and $g_d=3.8$ GeV$^{-2}$ in the framework of QCD sum rule.

For the parameter $\eta$, we use $\eta=0.4$ GeV$^{-2}$
\cite{1576}. The meson masses are from PDG: $m_{D^{0}}=1864.5$
MeV, $m_{D^{+}}=1869.3$ MeV. $m_{D_1}=2422.3$ MeV,
$m_{D^*{0}}=2006.7$ MeV, $m_{D^{*+}}=2010.0$ MeV, $m_{\pi}=135.0$
MeV, $m_{\rho}=775.5$ MeV, $m_{\omega}=782.7$ MeV \cite{PDG}.

\begin{center}
\begin{figure}[htb]
\begin{tabular}{c}
\scalebox{0.8}{\includegraphics{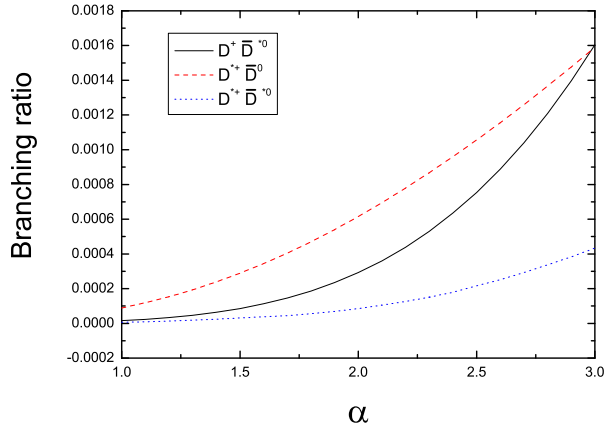}}
\end{tabular} \caption{The dependence of the branching ratios on $\alpha$. The solid,
dashed and dotted lines correspond to the $D^{+}\bar{D}^{*0}$,
$D^{*+}\bar{D}^{0}$ and $D^{*+}\bar{D}^{*0}$ modes respectively.
\label{decay}}
\end{figure}
\end{center}

With the above parameters, we plot the dependence of the branching
ratio of the two-body open charm decays $Z^+(4430)\to
D^{+}\bar{D}^{*0},D^{*+}\bar{D}^{0},D^{^*+}\bar{D}^{*0}$ on
$\alpha$ in Fig. \ref{decay}. In Table \ref{table}, we also list
show the branching ratios of $D^{+}\bar{D}^{*0}$,
$D^{*+}\bar{D}^{0}$, $D^{*+}\bar{D}^{*0}$ modes with typical
values of $\alpha$.

In short summary, we have calculated the two-body open charm decay
widths assuming $Z^+(4430)$ is a pseudoscalar molecular state. Our
numerical results show that the branching ratios of the two-body
decays $Z^+(4430)\to D^{+}\bar{D}^{*0}, D^{*+}\bar{D}^{0}
D^{*+}\bar{D}^{*0}$ are around $10^{-5}\sim 10^{-3}$, which is
strongly suppressed compared to the main decay modes $D^\ast {\bar
D}^\ast \pi$ if it's a molecular state with $J^P=0^-$. However,
they are only moderately suppressed compared with the discovery
mode $\psi'\pi^+$.

As an excellent candidate of the exotic states, the observation of
$Z^+(4430)$ has stimulated several speculations about its
structure. The tetraquark scheme seems an interesting choice.
However, there is no reliable dynamics forbidding the quarks and
anti-quarks to regroup and fall apart. With plentiful phase space
$Z^+(4430)$ will easily fall apart into final states such
$\psi\pi$, $\psi'\pi$, $\psi(1D)\pi$, $\eta_c\rho$,
$\chi_{cJ}\rho$, $D\bar D^\ast$ etc if $Z^+(4430)$ is a
tetraquark. Especially the two-body open charm modes
$D^{+}\bar{D}^{*0}, D^{*+}\bar{D}^{0}, D^{*+}\bar{D}^{*0}$ etc
will become one of the main decay modes, in sharp contrast with
the case of $Z^+(4430)$ being a molecular state. Thus these modes
provide a smoking gun to distinguish the molecular and tetraquark
schemes. We strongly urge experimentalists to look for these
two-body open charm decay modes together with the hidden charm
mode $\chi_{cJ}\rho^+$.

\begin{widetext}
\begin{center}
\begin{table}[th]
\caption{The branching ratios with typical values of
$\alpha$.\label{table}}
\begin{tabular}{c|cccccccccccccccccc}\hline\hline
$\alpha$&1.0&1.4&1.8&2.2&2.6&3.0\\\hline $B[Z^+\to
D^{+}\bar{D}^{*0}]$&$1.6\times 10^{-5}$&$6.4\times 10^{-5}$
&$1.9\times 10^{-4}$&$4.4\times 10^{-4}$
&$8.9\times 10^{-4}$&$1.6\times 10^{-3}$\\
$B[Z^+\to D^{*+}\bar{D}^{0}]$&$8.9\times 10^{-5}$&$2.4\times
10^{-4}$ &$4.7\times 10^{-4}$&$7.8\times 10^{-4}$
&$1.2\times 10^{-3}$&$1.6\times 10^{-3}$\\
$B[Z^+\to D^{*+}\bar{D}^{*0}]$& $6.3\times 10^{-6}$&$2.4\times
10^{-5}$ &$5.6\times 10^{-5}$&$1.3\times 10^{-4}$ &$2.5\times
10^{-4}$&$4.3\times 10^{-4}$
\\\hline\hline
\end{tabular}
\end{table}
\end{center}
\end{widetext}

\vfill

%%%%%%%%%%%%%%%%%%%%%%%%%%%%%%%%
\section*{Acknowledgments}
%%%%%%%%%%%%%%%%%%%%%%%%%%%%%%%%

This project was supported by the National Natural Science
Foundation of China under Grants 10625521, 10721063, 10705001 and
the China Postdoctoral Science foundation (20060400376). X.L. was
also supported by the \emph{Funda\c{c}\~{a}o para a Ci\^{e}ncia e
a Tecnologia of the Minist\'{e}rio da Ci\^{e}ncia, Tecnologia e
Ensino Superior} of Portugal (SFRH/BPD/34819/2007).

%\end{CJK}

\end{document}